\documentclass[aps,pre,twocolumn,showpacs,superscriptaddress,groupedaddress]{revtex4}  
\usepackage{graphicx}  
\usepackage{dcolumn}   
\usepackage{bm}        
\usepackage{amssymb}  
\usepackage{color} 
\begin{document}
\widetext
\title{Entropy and energy spectra in low-Prandtl-number convection with rotation
}
\author{Hirdesh K. Pharasi}
\author{Krishna Kumar}
\affiliation{Department of Physics, Indian Institute of Technology, Kharagpur-721 302, India}
{*e-mail: {\it kumar@phy.iitkgp.ernet.in}}
\author{Jayanta K. Bhattacharjee}
\affiliation{Harish-Chandra Research Institute, Allahabad-211 019,  India} 
\date{\today}

\begin{abstract}
We present results for entropy and kinetic energy spectra computed from direct numerical simulations for low-Prandtl-number ($Pr < 1$) turbulent flow in Rayleigh-B\'{e}nard convection with uniform rotation about a vertical axis. The simulations are performed in a three-dimensional periodic box for a range of Taylor number ($ 0 \leq Ta \leq 10^8$) and reduced Rayleigh number $r = Ra/Ra_{\circ} (Ta, Pr)$ ($1.0 \times 10^2 \le r \le 5.0 \times 10^3$). The Rossby number $Ro$ varies in the range $1.34 \le Ro \le 73$. The entropy spectrum $E_{\theta}(k)$ shows bi-splitting into two branches for lower values of wave number $k$. The entropy in the lower branch scales with $k$ as $k^{-1.4\pm 0.1}$ for  $r > 10^3$ for the rotation rates considered here. The entropy in the upper branch also shows scaling behavior with $k$, but the scaling exponent  decreases with increasing $Ta$ for all $r$. The energy spectrum $E_v(k)$ is also found to scale with the wave number $k$ as $k^{-1.4\pm 0.1}$ for $r > 10^3$. The scaling exponent for the energy spectrum and the lower branch of the entropy spectrum vary between $-1.7$ to $-2.4$ for lower values of $r$ ($< 10^3$). We also provide some simple arguments based on the variation of the Kolmogorov picture to support the results of simulations.

\end{abstract}
\maketitle
\section{Introduction}
Turbulence in Rayleigh-B\'{e}nard (RB) convection~\cite{siggia_1994, ahlers_review, lohse_xia_2010} has been studied extensively over the last three decades. There have been two primary objectives: \\
(A) to get a firm grasp on how the Nusselt number ($Nu$) scales with the Rayleigh number ($Ra$)~\cite{ahlers_review} and \\
(B) to understand the issues pertaining to energy and entropy spectra and fluxes~\cite{lohse_xia_2010}. \\
In the former~[i.e., (A)] the interest has centered on the scaling of $Nu$ with $Ra$ for very high values of $Ra$~\cite{kerr_1996, cioni_etal_1997, heslot_1987, castaing_1989, wu_etal_1990, niemela_etal_2000, grossmann_lohse_2000, kadanoff_2001, fauve_epl_2003} and on understanding the boundary layers which are responsible for the scaling and statistics of the temperature fluctuations. The study of RB convection under rotation started attracting a good deal of attention from the eighties and the effect of rotation on the Nusselt number has been repeatedly studied~\cite{niemela_etal_1986, julien_etal_1996, liu_ecke_1997, kunnen_etal_epl_2008, stevens_etal, king_etal_nature_2009, schmitz_tilgner_2010, pharasi_etal_pre_2011, stevens_2013}. 

In the latter [i.e., (B)] the interest has centered on whether the energy and entropy spectra will be determined by the Kolmogorov (K41)~\cite{K41} or the Bolgiano-Obukhov (BO) scaling~\cite{bolgiano_1959, obukhov_1959}. It is expected that there exists a crossover length $L_B$ (Bolgiano length) with corresponding wave number $k_B = 2\pi/L_B$ such that for wave number $k > k_B$ the energy spectrum is K41 ($k^{-5/3}$) and for $k < k_B$ it is BO ($k^{-11/5}$). While the outcomes of different experiments~\cite{ashkenazi_steinberg_1999, shang_xia_2001, zhou_xia_2001} are at best contentious, the numerical simulations~\cite{ calzavarini_etal_2002, kunnen_etal_pre_2008, mishra_etal_pre_2010}  show a scaling behavior which holds for much less than one decade of wave numbers. The effect of rotation on the energy and entropy spectra has never been considered. 

A re-examination of the K41 result for the RB convection with rotation (Sec. IV of this paper) showed that even for $k > k_B$ one  is not assured of Kolmogorov scaling. One has a lower limit $k_1$ and pure K41 can only be observed in the range $k_1 > k > k_B$. Similarly, one finds that there exists a wave number $k_2$ which limits the applicability of pure BO scaling to the range $k_B > k > k_2$. These restrictions are rather severe. The arguments indicate that one could at best expect an effective exponent. We decided to repeat the numerical simulations and try a scaling plot with no bias towards K41 or BO exponents. To our surprise the exponent that provided the best fit to the energy was $-1.5$. The error bars are small enough in  most of the runs to rule out $-5/3$. The K41 answer of $-5/3$ for the scaling exponent~\cite{kraichnan_1959, l'vov_1991, l'vov_1992, brandenburg_1992, mou_weichman_1993} relies on the fact that the sweeping of small eddies by large eddies is ignored, which is not a real part of turbulent motion. It is well known that the energy spectrum behaves as $k^{-3/2}$, if the dynamics is sweeping dominated. Straightforward arguments show that the scaling exponent would be $-5/2$ for the sweeping dominated BO regime. Hence it is clear that the numerical results presented in Sec. III support a sweeping dominated Kolmogorov regime. With this background, we decided to investigate the crossover between K41 and BO regimes by considering the Rayleigh-B\'{e}nard system under rotation. It should be possible to suppress the velocity fluctuations and give greater prominence to the thermal fluctuations in the rotation dominated regime, which in turn would be responsible for engineering a passage to the BO spectrum. Since the study of the spectrum for the RB turbulence has never been carried out for the rotating system, this could open up new possibilities.

\begin{table*}[tt]
\caption{\label{table1} List of the Prandtl number $Pr$, the Taylor number $Ta$, the Rayleigh number $Ra$, the reduced Rayleigh number $r = Ra/Ra_{\circ}$, the Rossby number $Ro = \sqrt{Ra/(Pr Ta)}$, the Nusselt number $Nu$, the dimensionless  Bolgiano wave number $k_{B} = 2\pi d/L_B$ corresponding to the  global Bolgiano length $L_B/d = \left(Nu \right )^{1/2}/{(Ra Pr)}^{1/4}$, and the the range of dimensionless wave numbers for scaling exponents $\gamma_1$, $\gamma_2$, and $\alpha$. The reduced Rayleigh number for the non-rotating case ($Ta = 0$) is defined as: $r = Ra/Ra_c$ with $Ra_c = 27 \pi^4/4$.}
\begin{ruledtabular}
\begin{tabular}{cccccccccc}
$Pr$&$Ta$&$Ra$&$r$&$Ro$&$Nu$&$k_{B}$ & \multicolumn{3}{c}{Range of $k$ for exponents} \\
\cline{8-10}
& & & & & & & $\gamma_1$ & $\gamma_2$ & $\alpha$ \\ \hline
$0.1$& $0$& $1.0\times10^5$& $1.5\times 10^2$& $\infty$& $5.05$& $27.96$& $6-25$& $4-23$& $7-21$ \\
$0.1$& $1.0\times10^{4}$& $1.76\times10^5$& $1.0\times10^2$& $13.27$& $6.10$& $29.30$& $6-25$& $5-26$& $8-29$\\
$0.5$& $1.0\times10^{4}$& $5.26\times10^5$& $1.0\times10^2$& $10.26$& $11.61$& $41.76$& $6-37$& $9-41$&$8-28$\\
$0.1$& $3.0\times10^{4}$& $2.24\times10^5$& $1.0\times10^2$& $8.64$& $6.67$& $29.76$& $6-31$& $4-24$&$9-31$ \\
$0.5$& $3.0\times10^{4}$& $8.90\times10^5$& $1.0\times10^2$& $7.70$& $13.74$&  $43.78$& $6-37$& $9-41$&$9-32$\\
$0.5$& $1.0\times10^{6}$& $6.83\times10^6$& $1.0\times10^2$& $3.70$& $21.23$& $58.62$& $6-50$& $9-60$& $9-56$ \\
$0.1$& $1.0\times10^{6}$& $1.07\times10^6$& $1.0\times10^2$& $3.27$& $9.11$& $37.65$& $6-37$& $9-35$& $8-39$ \\
$0.1$& $1.0\times10^{8}$& $8.97\times10^7$& $5.0\times10^2$& $2.99$& $28.95$& $63.91$& $6-62$& $28-95$& $30-160$ \\
$0.1$& $1.0\times10^{8}$& $1.79\times10^7$& $1.0\times10^2$& $1.34$& $15.80$& $57.82$& $12-75$& $29-57$& $29-75$ \\
$0.1$& $0$& $1.0\times10^6$& $1.5\times 10^3$& $\infty$& $9.51$& $36.23$& $6-37$& $4-30$& $4-42$ \\
$0.1$&$0$ & $2.0\times10^6$& $3.0\times 10^3$& $\infty$& $11.81$& $38.66$& $6-37$& $3-45$& $4-64$ \\
$0.1$&$1.0\times10^{4}$& $5.28\times10^6$ &$3.0\times 10^3$& $72.66$& $15.97$& $42.38$& $6-43$& $5-48$& $5-40$\\
$0.5$& $1.0\times10^{4}$& $1.58 \times 10^7$ & $3.0\times 10^3$& $56.21$& $30.51$& $60.31$& $6-56$& $8-89$& $6-60$\\
$0.1$&$3.0\times10^{4}$&$6.72\times10^6$&$3.0\times10^3$&$47.33$&$17.20$& $43.38$& $6-43$& $4-47$& $5-50$ \\
$0.5$& $3.0\times10^{4}$& $2.67\times10^7$& $3.0\times10^3$& $42.19$& $35.74$& $63.53$& $6-56$& $5-83$& $5-66$ \\
$0.1$& $1.0\times10^{6}$& $5.35\times10^7$& $5.0\times 10^3$& $23.13$& $31.72$& $53.70$& $6-56$& $8-72$& $8-98$ \\
$0.5$& $1.0\times10^{6}$& $2.05\times10^8$& $3.0\times 10^3$& $20.25$& $63.38$& $79.41$& $6-81$& $9-143$& $10-100$ \\
\end{tabular}
\end{ruledtabular}
\end{table*}

\begin{figure*}[ht]
\includegraphics[height=10 cm, width=17 cm]{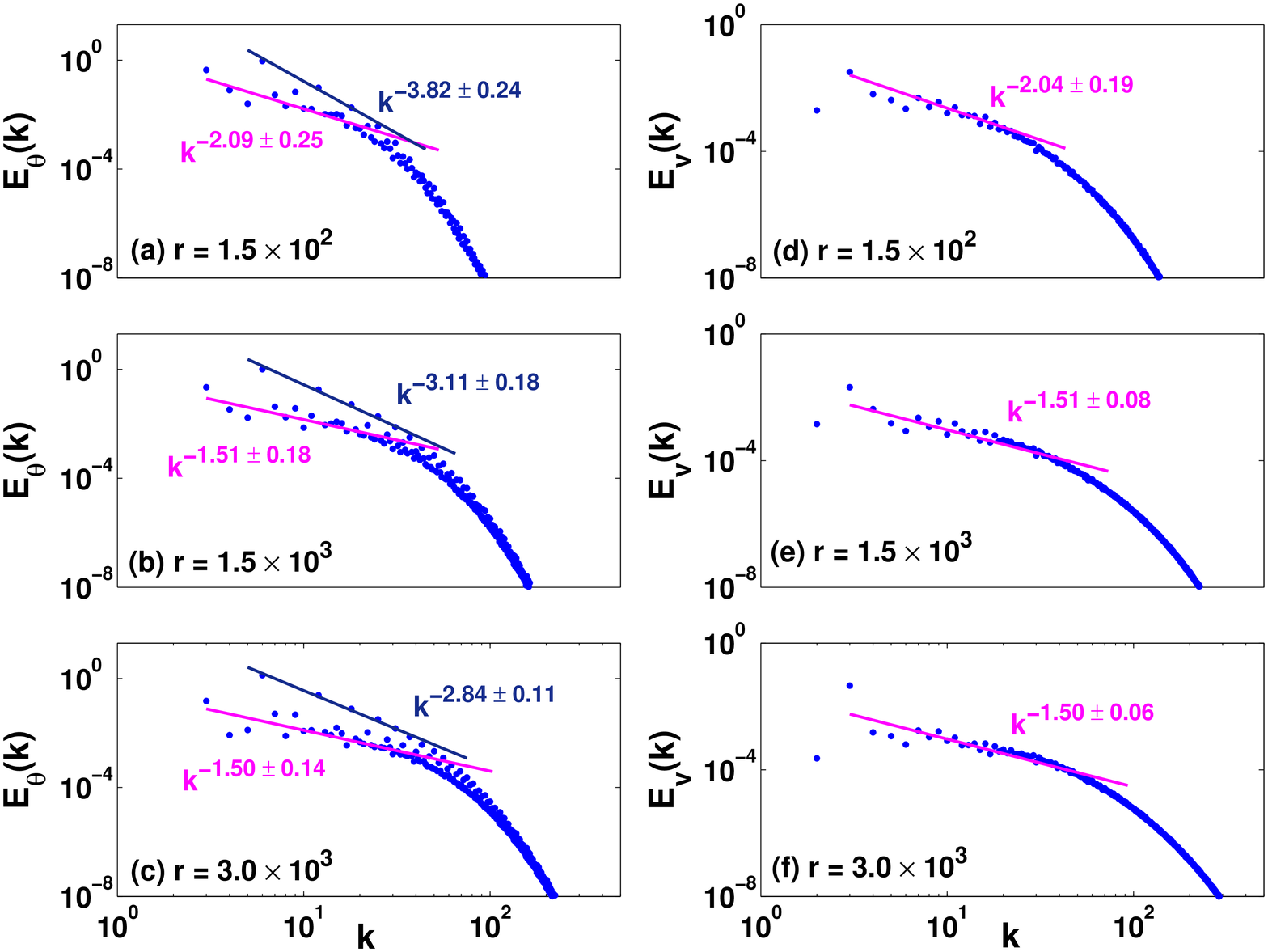}
\caption{(Color online) The entropy $E_{\theta}(k)$ (the left column) and the kinetic energy $E_v(k)$ (the right column) spectra obtained for $Pr=0.1$ and $Ta=0$ from the direct numerical simulations (DNS) on $512^3$ grid points for $r=1.5\times10^2$ [(a) $\&$ (d)], $r=1.5\times10^3$ [(b) $\&$ (e)], and $r = 3.0\times10^3$ [(c) $\&$ (f)]. The entropy spectrum shows bi-splitting. The best fit to the data points [blue (black) dots] corresponding to the upper and lower branches of the entropy spectrum are shown by the dark blue (black) and magenta (gray) lines respectively. The lower branch of the entropy spectrum $E_{\theta}(k)$ scales with $k$ as $k^{-1.5 \pm 0.1}$ for higher values of reduced Rayleigh number $r$ ($> 1.5\times10^3$). In the same range of $r$, the kinetic energy $E_v(k)$ also scales with wave number $k$ as $k^{-1.5 \pm 0.1}$. The best fit to the DNS data [blue (black) dots] for the energy spectrum is shown by the magenta (gray) line. For  $r = 1.5 \times 10^2$, the exponent is close to $-2$. The scaling exponent for the upper branch of the entropy spectrum varies with $r$.} \label{spectra_T0} 
\end{figure*}

For a RB system, the dynamics is controlled by two dimensionless numbers: the Rayleigh number $Ra$ and the Prandtl number $Pr$. The Rayleigh number is defined as $Ra = \alpha (\Delta T)g d^3/(\nu \kappa)$, where $\Delta T$ is the temperature difference between the bottom and top horizontal plates, which are separated by a distance $d$, $\alpha$ is the thermal expansion coefficient, $g$ the acceleration due to gravity, $\nu$ and $\kappa$ the kinematic viscosity and the thermal diffusivity respectively. The Prandtl number $Pr = \nu/\kappa$ is the ratio of the thermal diffusion time scale to the momentum diffusion time scale. Convection sets in as $Ra$ is raised above a critical value $Ra_c$, which is $27 \pi^4/4$ for {\it stress-free} boundaries and 1708 for {\it no-slip} boundaries. The threshold is independent of $Pr$. Convective turbulence occurs for $Ra \gg Ra_c$. It is convenient to describe turbulence in terms of the reduced Rayleigh number $r = Ra/Ra_c$. The convection appears at $r = 1$, and the convective flow becomes turbulent for $r \gg 1$. 

There is another dimensionless number in the presence of uniform rotation about a vertical axis: the Taylor number $Ta = 4 \Omega^2 d^4/\nu^2$, where $\Omega$ is the rotation frequency. The threshold is now strongly dependent on $Ta$. For $Pr > 0.667$ the onset of convection is always stationary for all values of $Ta < Ta_c (Pr)$, where $Ta_c (Pr)$ is the threshold of K\"{u}ppers-Lortz instability. The threshold for stationary convection  $Ra_c (Ta)$ is proportional to  $Ta^{2/3}$ for $1 \ll Ta <Ta_c (Pr)$. For $Pr < 0.667$, the onset is oscillatory and the threshold $Ra_{\circ} (Ta, Pr)$ depends on $Ta$ as well as $Pr$. For very small values of $Pr$, the oscillatory threshold $Ra_{\circ}$ is proportional to $Pr^{4/3} Ta^{2/3}$ as $Ta \gg 1$. The importance of the effect of rotation can be determined from another dimensionless parameter which is expressed as $Ro = \sqrt{Ra/(Pr Ta)}$ and is called the Rossby number. Generally a Rossby number smaller than unity or closer to unity corresponds to the dominance of rotation. It is important to discuss the known results about the Nusselt number investigations in both non rotating~\cite{ castaing_1989, wu_etal_1990, niemela_etal_2000, grossmann_lohse_2000, kadanoff_2001} and rotating~\cite{king_etal_nature_2009, schmitz_tilgner_2010, pharasi_etal_pre_2011} systems. The Nusselt number, unlike the spectrum, has been studied extensively and as we shall see has some bearing on our findings about the spectrum. It is established that in the non rotating systems, the Nusselt number follows a slightly modified power law. In the early work of Castaing et al.~\cite{castaing_1989}, physical arguments and experimental data were presented to support a simple power law $Nu  \propto  Ra^{2/7}$. A decade later Niemela et al.~\cite{niemela_etal_2000} suggested an exponent of 0.3 with a logarithimic correction and Grossmann and Lohse~\cite{grossmann_lohse_2000} proposed a sum of two terms with exponents $1/3$ and $1/4$. The later proposals were  refinements on the original proposal of Castaing et al.~\cite{castaing_1989} which, in the range of Rayleigh numbers that we will be discussing, is an adequate representation. This form of $Nu$ vs $Ra$ curves were found to be true for a wide range of Prandtl numbers. 

The effect of rotation on the Nusselt number has also been studied extensively as noted earlier but it is only lately that the effect of rotation has been cast in a particularly useful form. Beginning with the work of King et al.~\cite{king_etal_nature_2009}, we see that at a given rotation speed, the Nusselt number follows the curve for the non-rotating situation for high Rayleigh numbers (see also,~\cite{schmitz_tilgner_2010, pharasi_etal_pre_2011}) and as the Rayleigh number is lowered, the Nusselt number falls below the non rotating value at a particular value of Ra which we call $Ra_t$.  As the rotation speed is changed the value of $Ra_t$ changes - decreasing as the rotation speed decreases. This is a common feature of all the data taken at different Prandtl numbers and Taylor numbers. We decided to check whether there is a systematic way of characterizing $Ra_t (Ta, Pr)$. We found that if the threshold of convection is $Ra_0 (Ta, Pr)$, then the ratio  $r_t = Ra_t(Ta, Pr) / Ra_0(Ta, Pr)$ depend on $Pr$ and $Ta$. For $r_t$ is approximately $1.0 \times 10^2$ for $Ta = 10^6$ and $Pr = 0.1$. For a given rotation rate, $r_t$ is lower for higher values of $Pr$. Sec III we will see that it has interesting connections with the scaling in the spectrum.

The direct numerical simulations of Sec. III are carried out for a wide range of dimensionless parameters on $256^3$ and $512^3$ grids. The Taylor number  is varied in the range $ 0 \leq Ta \leq 10^8$. The onset of convection is always oscillatory for the values of $Pr$ and $Ta$ considered here. The reduced Rayleigh number is therefore defined as $r = Ra/Ra_{\circ} (Ta, Pr)$, where $Ra_{\circ}(Ta,Pr)$ is the threshold value for the onset of oscillatory convection. The reduced Rayleigh number is varied in the range $10^2 \leq r  \leq 5.0 \times 10^3$. This allowed a variation of the Rossby number in a range ($1.34 \le Ro \le 73$) for the rotating convection. The entropy spectrum $E_{\theta}(k)$ shows two branches at smaller values of $k$, both of which show scaling behavior. The lower branch of the entropy spectrum scales with $k$ as $k^{-1.4 \pm 0.1}$ for higher values of $r$ ($10^3 \le r \le 5.0 \times 10^3$) for all values of $Ta$. The scaling exponent is universal, and is observed for wave numbers much beyond $k_B$. The scaling exponent for the upper branch of the entropy spectrum is found to vary between $-2$ and $-4$. The scaling exponent of the energy spectrum is found close to $-1.5$ for $10^3 \le r \le 5.0 \times 10^3$. The exponents of the energy and the entropy spectra for lower values of $r$ ($10^2 \le r < 10^3$) are found to lie between  $-1.7$ and $-2.4$ showing that the flow has not entered an asymptotic regime either for Kolmogorov or Bolgiano scaling, in agreement with the arguments presented in Sec. IV. 
\begin{figure*}[ht]
\includegraphics[height=10 cm, width=17 cm]{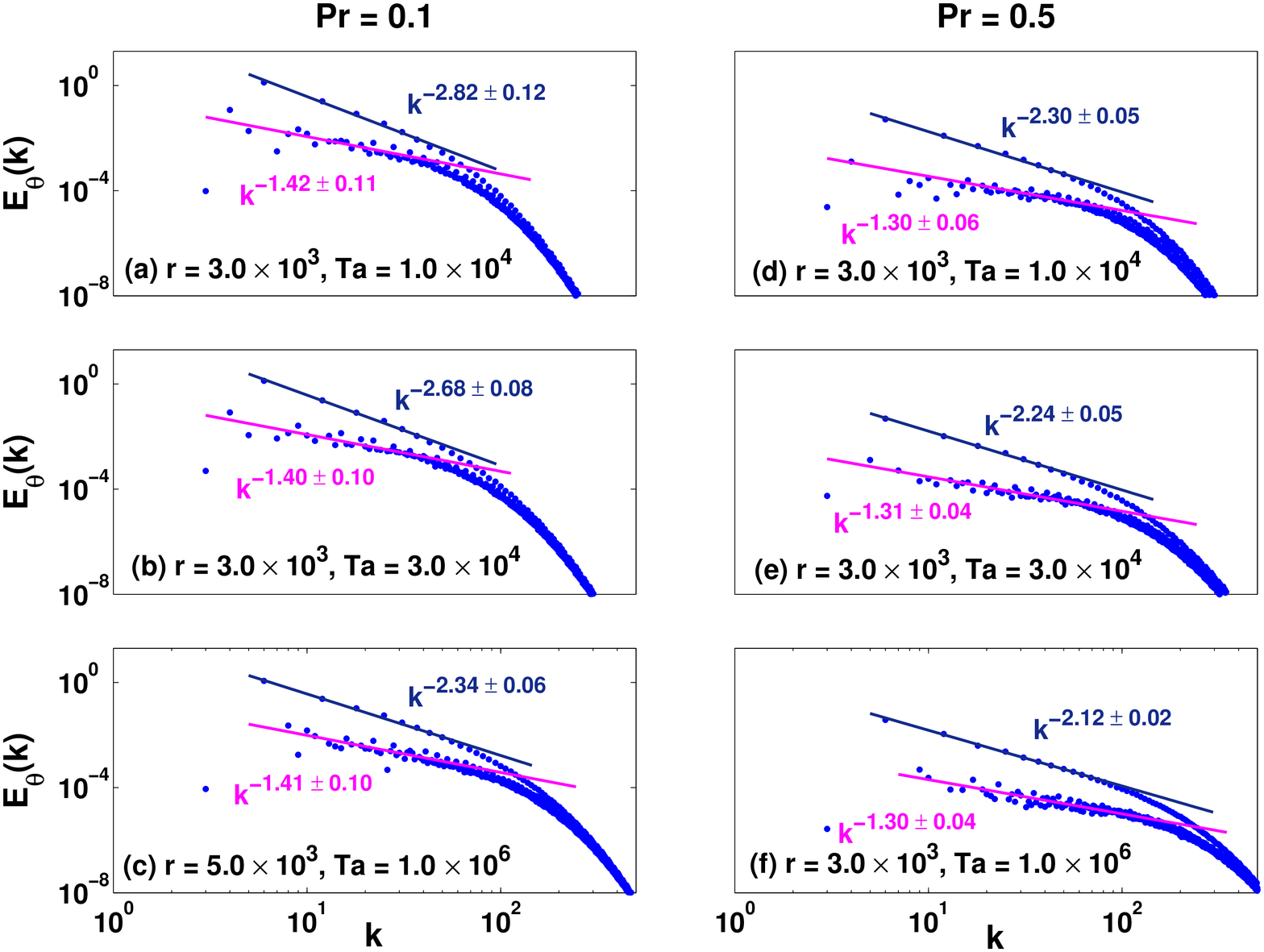}
\caption{(Color online) The entropy spectrum $E_{\theta}(k)$ for $Pr = 0.1$ (the left column) and $Pr = 0.5$ (the right column) computed from DNS on $512^3$ grid points for (a) $r = 3.0 \times 10^3$ \& $Ta = 1.0 \times 10^4$, (b) $r = 3.0 \times 10^3$ \& $Ta = 3.0 \times 10^4$, (c)  $r = 5.0 \times 10^3$ \& $Ta = 1.0 \times 10^6$ (d) $r = 3.0 \times 10^3$ \& $Ta = 1.0 \times 10^4$, (e) $r = 3.0 \times 10^3$ \& $Ta = 3.0 \times 10^4$, and (f) $r = 3.0 \times 10^3$ \& $Ta = 1.0 \times 10^6$. The best fit to the data points [blue (black) dots] in the upper and the lower branches shown by the dark blue (black) and magenta (gray) lines respectively. The upper branch shows scaling trends: $E_{\theta}(k) \sim k^{-\gamma_1}$ with the scaling exponent $-\gamma_1$ lying between $-2.1$ and $-2.9$. The lower branch of the entropy spectrum $E_{\theta}(k)$ scales with  $k$ as $k^{-\gamma_2}$ with $\gamma_2 = 1.4 \pm 0.1$ for $Pr = 0.1$ and $1.3 \pm 0.1$ for $Pr = 0.5$.} \label{entropy_spectrum1} 
\end{figure*}

\begin{figure}[h]
\includegraphics[height=8.5 cm, width=8.5 cm]{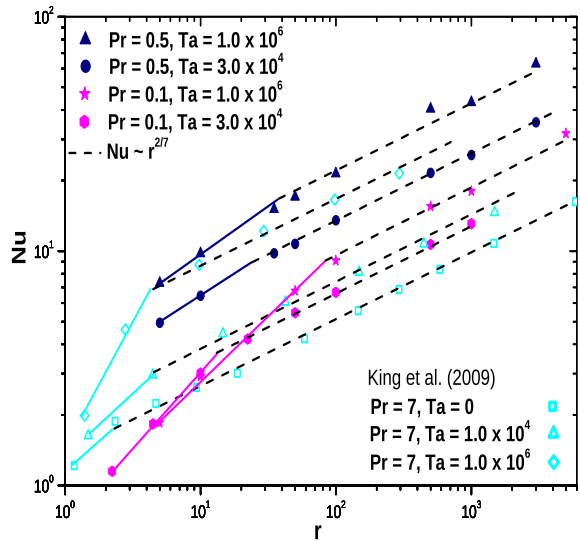}
\caption{(Color online)  Variation of the Nusselt number $Nu$ with the reduced Rayleigh number $r$ for different values of Prandtl number $Pr$ and Taylor number $Ta$. Data points in blue (black), magenta (gray) and cyan   (light gray) colors are for $Pr = 0.1$, $0.5$ and $7$ respectively. Data points for $Pr = 7$ are taken from King et al.~\cite{king_etal_nature_2009}. The dashed black lines describe the scaling $Nu \propto r^{2/7}$ for different values of $Pr$ and $Ta$. The exponent $2/7$ is almost independent of $Pr$ and $Ta$ for $r > r_t (Ta, Pr)$.} \label{nu_r} 
\end{figure}

\begin{figure*}[ht]
\includegraphics[height=10 cm, width=17 cm]{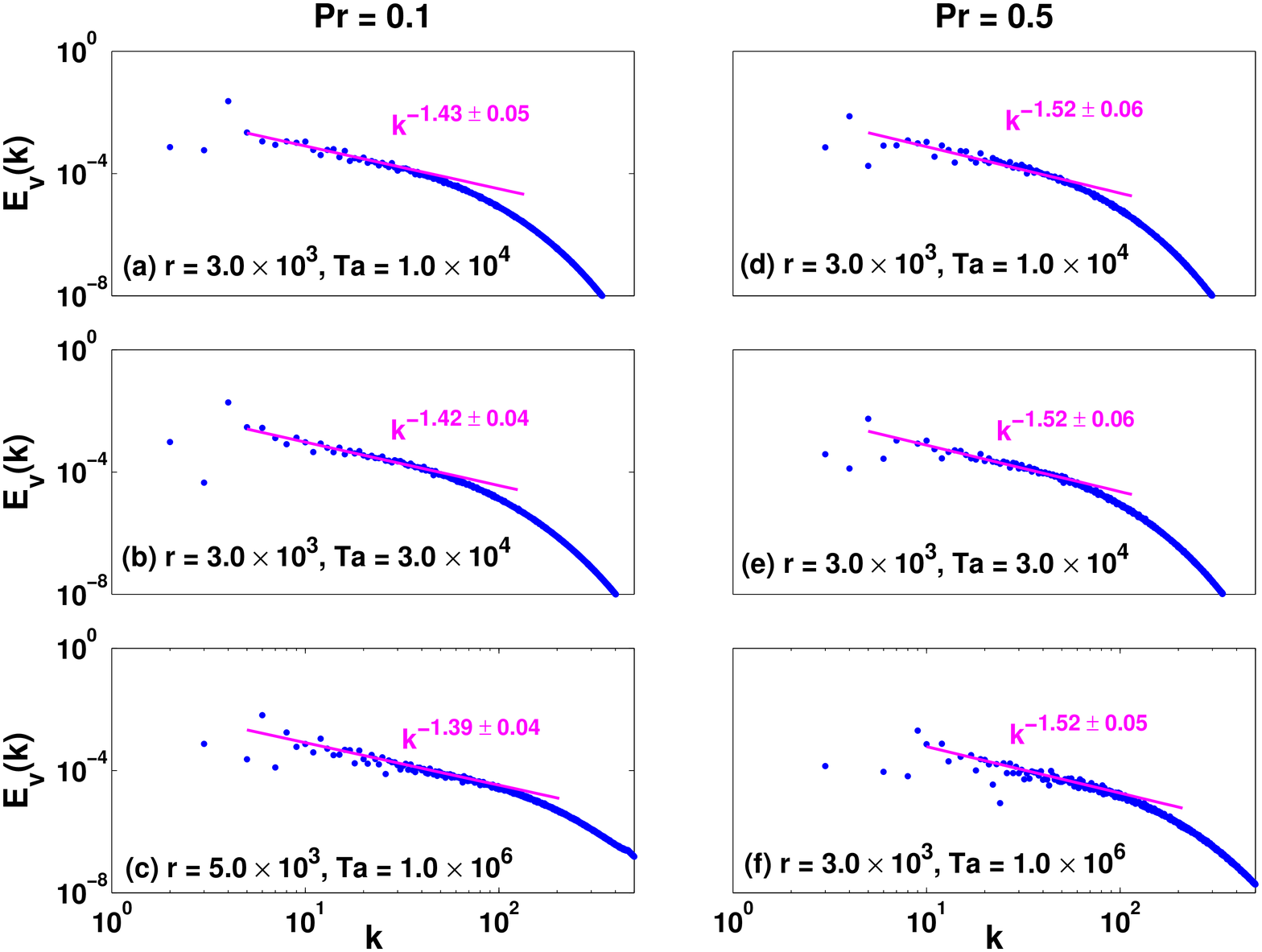}
\caption{(Color online) The energy spectrum $E_v(k) $ for $Pr = 0.1$ (the left column) and $Pr = 0.5$ (the right column) computed from DNS for (a) $r = 3.0 \times 10^3$ \& $Ta = 1.0 \times 10^4$, (b) $r = 3.0 \times 10^3$ \& $Ta = 3.0 \times 10^4$, (c)  $r = 5.0 \times 10^3$ \& $Ta = 1.0 \times 10^6$ (d) $r = 3.0 \times 10^3$ \& $Ta = 1.0 \times 10^4$, (e) $r = 3.0 \times 10^3$ \& $Ta = 3.0 \times 10^4$, and (f) $r = 3.0 \times 10^3$ \& $Ta = 1.0 \times 10^6$. The magenta (gray) line is the best fit to DNS data points [blue (black) dots]. The kinetic energy $E_v(k)$ scales with the wave number $k$ as $k^{-\alpha}$. The value of the scaling exponent $\alpha = 1.4 \pm 0.1$ for $Pr = 0.1$ and $\alpha = 1.5 \pm 0.1$ for $Pr = 0.5$.} \label{energy_spectrum1}
\end{figure*}

\section{The hydrodynamic system and direct numerical simulations}
The fluid thickness $d$, $\sqrt{\alpha ({\Delta T}) g d}$ and $\nu ({\Delta T})/\kappa$ are used to make all lengths, the velocity field $\mathbf{v} (x,y,z,t)$, and the temperature field $\theta (x,y,z,t)$ respectively dimensionless. The equations of motion, in Boussinesq approximation, then read as:
\begin{equation}
\partial_t \mathbf{v} + (\mathbf{v}\cdot\boldsymbol{\nabla})\mathbf{v} = -\boldsymbol{\nabla}p + Pr\theta \hat{\bf{z}} + {\sqrt{\frac{Pr}{Ra}}} \nabla^{2}\mathbf{v}-{Ro^{-1}} (\hat{\bf{z}}\mathbf{\times}\mathbf{v}),\label{momentum}
\end{equation}
\begin{equation}
Pr\left(\partial_t \theta + \mathbf{v}\cdot\boldsymbol{\nabla}\theta\right)  = {\sqrt{\frac{Pr}{Ra}}} \nabla^{2}\theta + w, \label{heat} \label{temperature}
\end{equation}
\begin{equation}
\boldsymbol{\nabla}\cdot\mathbf{v} = 0,\label{continuity}
\end{equation}
where $p(x,y,z,t)$ the pressure field, and the symbol  $\hat{\bf z}$ stands for a unit vector directed vertically upward. We have considered thermally conducting and {\it stress-free} top and bottom boundaries. This leads to the boundary conditions: $\partial_{z}v_1 = \partial_{z}v_2 = v_3 = \theta =0$ at $z=0$ and $z=1$. All the fields are considered periodic in horizontal plane.   
The velocity, temperature and pressure fields are expanded as:
\begin{equation}
v_1 (x,y,z,t) = \sum_{l,m,n} U_{lmn}(t) e^{ik_o(lx+my)} \cos{(n\pi z)},\label{eq.u}
\end{equation}
\begin{equation}
v_2 (x,y,z,t) = \sum_{l,m,n} V_{lmn}(t) e^{ik_o(lx+my)} \cos{(n\pi z)}, \label{eq.v}
\end{equation}
\begin{equation}
v_3 (x,y,z,t) = \sum_{l,m,n} W_{lmn}(t) e^{ik_o(lx+my)} \sin{(n\pi z)},\label{eq.w}
\end{equation}
\begin{equation}
\theta(x,y,z,t) = \sum_{l,m,n} \Theta_{lmn}(t) e^{ik_o(lx+my)} \sin{(n\pi z)},\label{eq.th}
\end{equation}
\begin{equation}
p (x,y,z,t) = \sum_{l,m,n} P_{lmn}(t) e^{ik_o(lx+my)} \cos{(n\pi z)},\label{eq.p}
\end{equation}
where $U_{lmn}(t)$, $V_{lmn}(t)$, $W_{lmn}(t)$,  $\theta_{lmn}(t)$, and $P_{lmn}(t)$ are the Fourier amplitudes in the expansion of the fields $v_{1}$, $v_{2}$, $v_{3}$, $\theta$, and $p$ respectively. The integers $l,m,n$ can take values consistent with the continuity equation~(Eq.~\ref{continuity}). The difference $Ra - Ra_{\circ}(Ta,Pr)$ decreases with increase in $Ta$. The Rayleigh number which for $Ta = 0$ corresponded to $r \gg 1$ is no longer in that regime for $Ta \gg 1$, and we need to go to even larger $Ra$ to achieve $r \gg 1$.  The reduced Rayleigh number $r$ instead of actual Rayleigh number $Ra$ is therefore a more appropriate parameter to describe turbulent regimes in rotating Rayleigh-B\'{e}nard convection. As the primary convection is oscillatory for the fluids we have investigated, we set $r = Ra/Ra_{\circ}$ in the presence of rotation. However, $r = Ra/Ra_c$ with $Ra_c = 27 \pi^4/4$ is used in the absence of rotation ($Ta =0$). The critical Rayleigh number $Ra_{\circ} (Ta, Pr)$ and the corresponding wave number $k_{\circ} (Ta, Pr)$ at the onset of oscillatory convection~\cite{chandrasekhar:book_1961} with stress-free boundaries are given as:
\begin{equation}
Ra_{\circ} (Ta, Pr) = 2\left( \frac{1 + Pr}{k^2_{\circ}}\right) \left[ (\pi^2 + k^2_{\circ})^3 + \frac{\pi^2 Ta Pr^2}{(1 + Pr)^2}\right],
\end{equation}
where $k_{\circ}$ is a real positive solution of the equation:
\begin{equation}
2\left( \frac{k_{\circ}^2}{\pi^2}\right)^3+3\left(\frac{k_{\circ}^2}{\pi^2}\right)^2 = \left[1+\frac{Pr^2}{(1+Pr)^2}\frac{Ta}{\pi^4}\right].
\end{equation}

\begin{figure*}[ht]
\includegraphics[height=10 cm, width=17 cm]{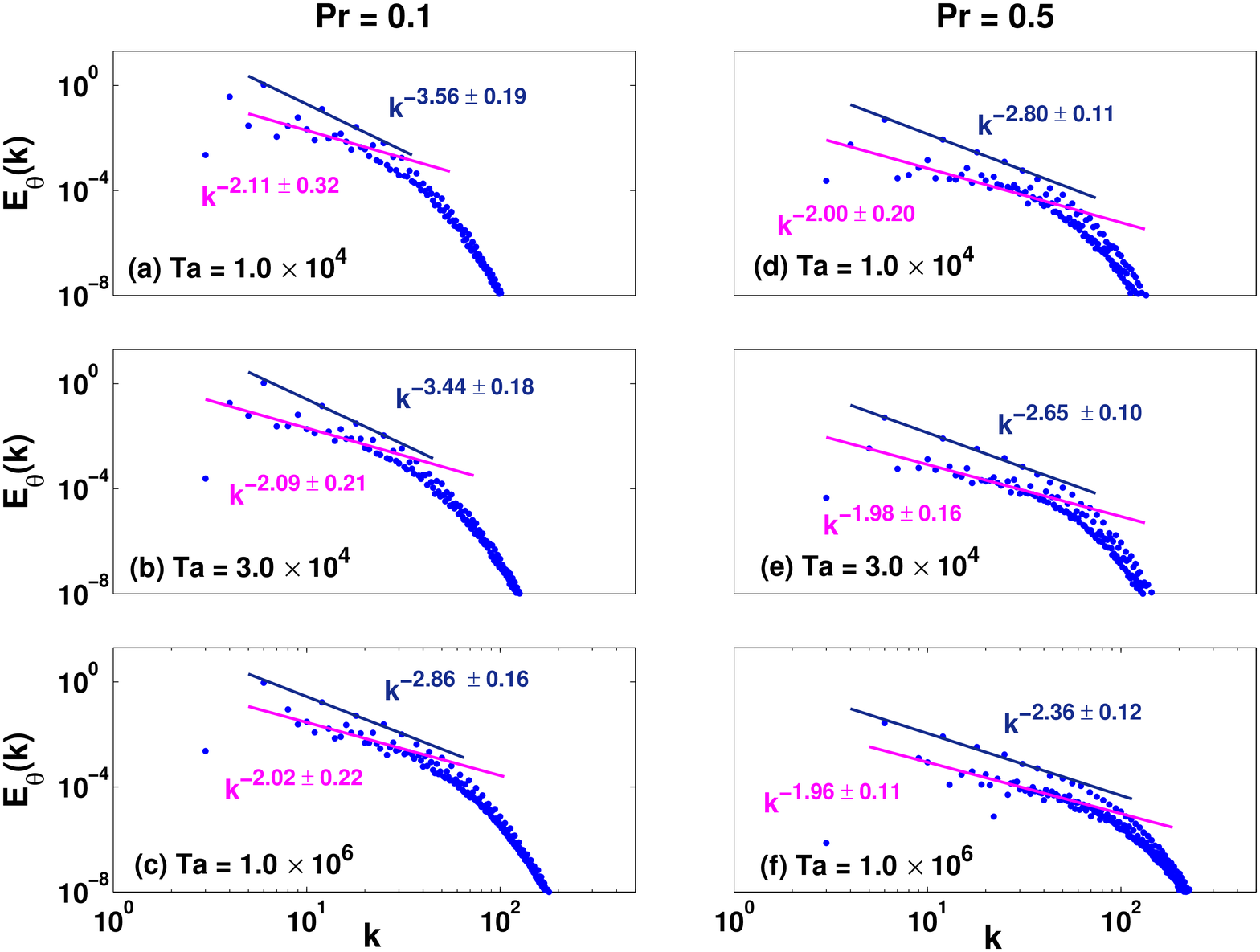}
\caption{(Color online) The entropy spectrum $E_{\theta}(k)$ [blue (black) dots] for $Pr = 0.1$ (the left column) and for $Pr = 0.5$ (the right column) computed at $r = 1.0 \times 10^2$. The plots are for $Ta = 1.0 \times 10^4$ [(a) \& (d)], $Ta = 3.0 \times 10^4$ [(b) \& (e)], and $Ta = 1.0 \times 10^6$ [(c) \& (f)] respectively. The lower branch of entropy spectrum $E_\theta(k)$ scales with wave number $k$ as $k^{-\gamma_2}$ with $\gamma_2 = 2.1 \pm 0.3$. The upper branch also shows scaling behavior with the scaling exponent lying between $-2.2$ and $-3.8$. The color code is the same as used in Fig.~\ref{entropy_spectrum1}.} \label{entropy_spectrum2} 
\end{figure*}

\begin{figure*}[ht]
\includegraphics[height=10 cm, width=17 cm]{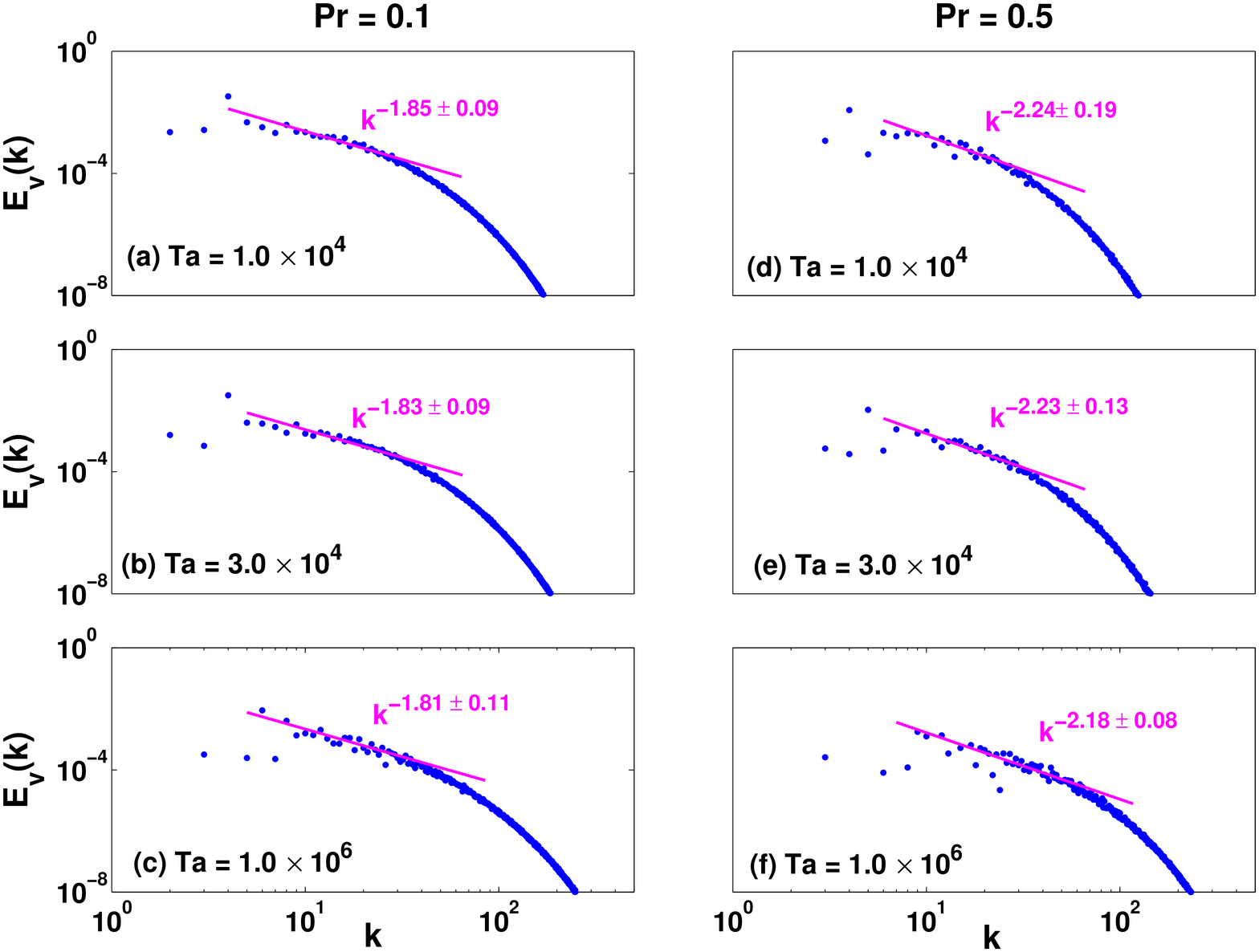}
\caption{(Color online) The energy spectrum $E_v(k) $ for $Pr = 0.1$ (the left column) and $Pr = 0.5$ (the right column) computed from DNS at $r = 1.0 \times 10^2$. The spectra are for $Ta = 1.0 \times 10^4$ [(a) \& (d)], $Ta = 3.0 \times 10^4$ [(b) \& (e)], and $Ta = 1.0 \times 10^6$ [(c) \& (f)] respectively.  The kinetic energy $E_v(k)$ scales with wave number $k$ as $k^{-\alpha}$ with $\alpha = 1.8 \pm 0.1$ for $Pr = 0.1 $ and $\alpha = 2.2 \pm 0.2$ for $Pr = 0.5$. The color code is the same as used in Fig.~\ref{energy_spectrum1}.} \label{energy_spectrum2} 
\end{figure*}

\section{Entropy and energy spectra}
We have numerically integrated the hydrodynamic system (Eqs.~(\ref{momentum})-(\ref{continuity})) with stress-free boundary conditions using an open-source code TARANG~\cite{tarang} based on pseudo spectral method. All the simulations are done in a box of size $L_x\times L_y\times 1$, where $L_x = L_y = 2\pi/k_o(Ta,Pr)$ for the purpose. The fourth order Runge-Kutta (RK4) scheme is used for the time advancement. The time steps have been monitored to have CFL condition satisfied all the time. There are two dissipative (Kolmogorov) scales: (i) The minimum dissipative scale for the  kinetic energy is $\eta^{K}_{min}=\left( \nu^{3}/\epsilon_{max}\right)^{1/4}$, where $\epsilon^{K}_{max} = \left[ \nu \left( \partial_{i} v_{j} (x,y,z,t) \right)^2\right] _{max}$ is the maximum dissipation rate of the kinetic energy, and (ii) the minimum dissipative scale for ``thermal energy'' is $\eta^{\Phi}_{min}=\left( \kappa^{3} / \epsilon_{max} \right)^{1/4}$, where $\epsilon^{\Phi}_{max} = \left[ \kappa\left(\partial_{i} \theta(x,y,z,t) \right)^2 \right] _{max}$ is the  maximum dissipation rate of ``thermal energy''. The grid-size has to be chosen such that the smallest dissipative (Kolmogorov) scale is resolved. The grid size $l_{grid}$ should be smaller than the lower value of  $\eta^{K}_{min}$ and $\eta^{\Phi}_{min}$. In practice, one uses the box averaged mean dissipation rate $\langle \eta \rangle = \left[ \nu^{3}/\langle \epsilon \rangle \right]^{1/4}$. That is, $l_{grid} < \langle \eta \rangle$. This leads to a grid size based cut-off wave number $k_{max} \approx 1/ l_{grid}$. This yields the condition for grid resolution: $k_{max} \times \langle \eta \rangle > 1$.  For several computational works on convective turbulence, this product lies in the range of $1 < k_{max} \times \langle \eta \rangle < 2$ (e.g.,~\cite{kaneda_etal_2003}). The grid resolutions for all the simulations presented here are such that the product lies in the range $2 <  k_{max} \times \langle \eta \rangle \le 15$, which is good enough to resolve the dissipative scales for all values of $Ta$, $r$ and $Ro$ considered here. The hydrodynamic equations were first integrated for approximately $100$ dimensionless time units on $256^3$ grids for this purpose. The final values of all the fields were then used to continue a simulation on $512^3$ grids. 

Table~\ref{table1} gives the details of simulations.  It also lists the Rossby numbers $Ro$, the Nusselt number $Nu$, the global dimensionless Bolgiano length $L_B/d = \left(\overline{Nu} \right )^{1/2}/{(Ra Pr)}^{1/4}$ and the corresponding cut-off wave number $k_{B} = 2\pi/L_B$ for different values of  $Ta$ and $r$. The over-line stands for the time average. The global Bolgiano length has been computed by taking the time average of $Nu$, which is already a spatially (box) averaged quantity. A very long signal for $Nu$ is used for the purpose of the time averaging, as only one number is required to be stored at each instant of time. The global Bolgiano length decreases with increasing value of $r$ for a fixed values of $Ta$. The dimensionless local Bolgiano length has been computed by the first principle. The velocity and the temperature fields are first computed at all grid points. The gradients of these fields were then computed on these grid points. The time average of the horizontally averaged fields on all grid points are then used for the determination of the local Bolgiano length. The local Bolgiano length $l_B/d$ shows almost a constant value in the central part of the fluid layer. The global Bolgiano length is found to be one order magnitude higher than the local Bolgiano length in the central part of the simulation box.  Kunnen et al.~\cite{kunnen_etal_pre_2008} also observed a difference of one order of magnitude in the local and the global Bolgiano lengths. As $L_B$ and $l_B$ are computed by different averaging procedures, they are two different quantities. The computation of $l_B$ requires storing $512^3$ numbers for each of the $v_1$, $v_2$, $v_3$ and $\theta$ fields at every instant of time for $512^3$ grid points. The huge data set limits the computation and the storage of all fields for shorter time. This may add, on time averaging, further difference in the two quantities. All the scaling exponents discussed in this paper are computed from using the global quantities. 

The wave number space is divided into several spherical shells. The symbols $E_\theta(k_i)$ and $E_v(k_i)$ represent the entropy and the energy respectively in the $i$th spherical shell of inner radius $k_i$ and outer radius $k_{i+1}$, where $i$ is an integer. The entropy spectrum $E_\theta(k_i)$ and the energy spectrum $E_v(k_i)$ are defined as:
\begin{eqnarray}
E_{\theta}(k_i)&=& \sum_{k_i {\leqslant k} < k_{i+1}} \frac{1}{2} |\theta({\bf k})|^2\\
E_v(k_i)&=& \sum_{k_i {\leqslant k} < k_{i+1}} \frac{1}{2} |v({\bf k})|^2.
\end{eqnarray}
The wave number $k$ is given by,
\begin{equation}
k=\left[ k_o^2 (l^2 + m^2) + \pi^2 n^2\right]^{1/2}.
\end{equation}
Therefore, the data points computed from DNS appear only at $k = k_i$ in all the spectra $E_{\theta} (k)$ and $E_v (k)$. 

\begin{figure*}[ht]
\includegraphics[height=7 cm, width=17 cm]{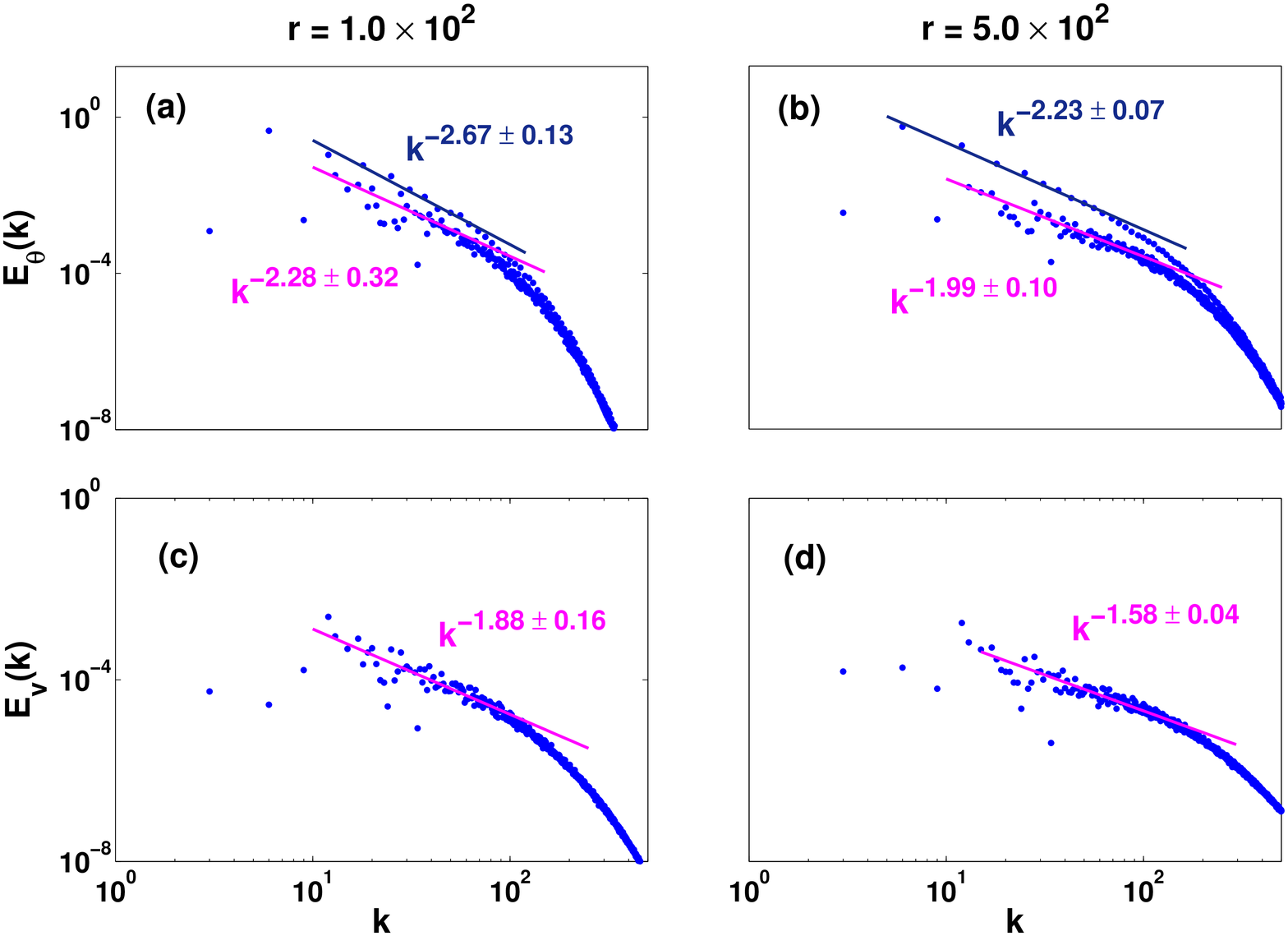}
\caption{(Color online) The entropy (the upper row) and the energy (the lower row) spectra, as computed from DNS for $Pr = 0.1$ and $Ta = 1.0 \times 10^8$. The entropy spectra $E_{\theta}(k)$ for (a) $r = 1.0 \times 10^2$ and (b) $r = 5.0 \times 10^2$. The energy spectra $E_v(k)$ for (c) $r = 1.0 \times 10^2$ and (d) $r = 5.0 \times 10^2$. The energy $E_v(k)$ scales with wave number $k$ as $k^{-\alpha}$ with $\alpha = 1.75 \pm 0.25$. The color code is the same as used in Fig.~\ref{spectra_T0}.} \label{spectrum_high_T} 
\end{figure*}

\begin{figure}[b]
\includegraphics[height=12 cm, width=8.5 cm]{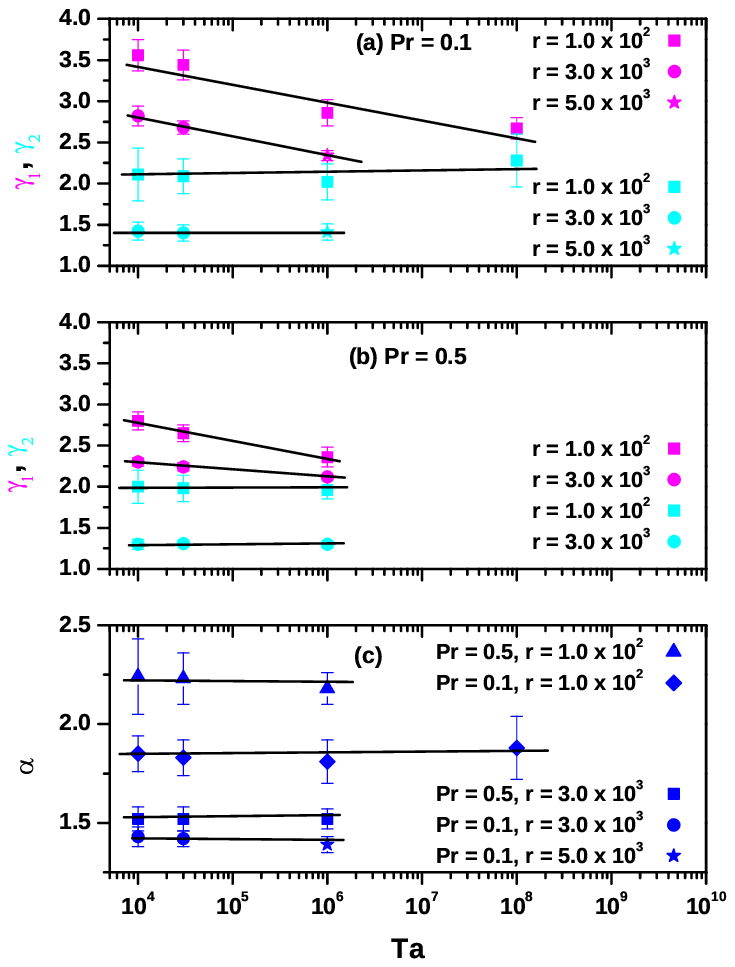}
\caption{(Color online) The variation of scaling exponents for the entropy and the energy spectra with $Ta$ for different values of $r$ and $Pr$. The variation of the exponents $\gamma_1$ [in magenta (gray) color] and $\gamma_2$ [in cyan (light gray) color] with $Ta$ for (a) $Pr = 0.1$ and (b) $Pr = 0.5$.  (c) The variation of the  exponent $\alpha$ with $Ta$ [in blue (black) color] for $Pr = 0.1$ and $Pr =0.5$.} \label{exonents} 
\end{figure}

Figure~\ref{spectra_T0} displays the entropy and the energy spectra $E_{\theta} (k)$ and $E_v (k)$ respectively for a fluid of $Pr = 0.1$ at different values of $r$ in the absence of rotation ($Ta = 0$), which corresponds to $Ro \rightarrow \infty$ situation.  The entropy spectrum (the left column) shows bi-splitting, which is consistent with earlier observations~\cite{mishra_etal_pre_2010, vincent_1999}. The expansion of convective temperature has two types of Fourier modes: the modes $\theta (0,0,n)$ which depend on the vertical coordinates only and modes $\theta (l,m,n)$ which depend on the horizontal as well as the vertical coordinates. The modes $\theta (0,0,n)$ lead to a nonzero value of the horizontally averaged convective temperature field ($<\theta>_H$). These modes therefore contribute to the thermal flux across the fluid layer. The upper branch of the spectrum is due to modes $(0,0,2n)$. The lower branch of the entropy spectrum is due to the Fourier modes which depend on both the horizontal and vertical coordinates. Their horizontal average vanishes, i.e., $<\theta>_H = 0$.  We believe that the bi-splitting of the entropy is independent of the velocity boundary conditions. The two kinds of modes for the temperature field always exist for both no-slip and free-slip velocity boundary conditions. Most of the experiments reported the entropy spectra in the frequency space instead of wave number space. This may be a possible reason for not observing bi-splitting in experiments. Both branches of the entropy spectrum show scaling behavior. We determined the scaling exponents by doing the best fit to the data obtained from DNS. The upper branch of the entropy spectrum was found to scale with the wave number $k$ as $k^{-\gamma_1}$. The exponent $\gamma_1$ was found close to $3$ for $r > 1.5 \times 10^3$ and close to $3.8$ for smaller values of $r$ ($< 10^3$). The lower branch of the entropy spectrum varied with $k$ as $k^{-\gamma_2}$. The exponent $\gamma_2$ was found closer to $2$ for lower values of $r$ ($10^2 < r < 10^3$) and $3/2$ for higher values of the $r$ ($> 1.5 \times 10^3$). The bi-splitting point and the scaling regime have been found to shift towards higher values of $k$ for higher values of $r$. One assumes $<\theta>_H = 0$ to predict scaling exponent for the entropy spectrum using phenomenological arguments. The scaling behavior for the lower branch of the entropy spectrum should therefore be used for any comparison with such theoretical predictions. The right column of Fig.~\ref{spectra_T0} shows the kinetic energy spectra for different values of $r$ for non-rotating case. The energy has been found to scale with $k$ as $k^{-\alpha}$. The best fit determined the scaling exponent $\alpha = 3/2$ for $r > 1.5 \times 10^3$. Its value was found equal to $2 \pm 0.2$ for $r < 10^3$.

Figure~\ref{entropy_spectrum1} displays the entropy spectra for higher values of $r$ ($> 1.5 \times 10^3$) at different Taylor numbers $Ta$. The left column shows the spectra for $Pr = 0.1$ and the right column for $Pr = 0.5$. The entropy spectrum is found to show bi-splitting, as in the absence of rotation. Both the branches of entropy spectrum show scaling behavior. The scaling exponent $\gamma_2$ for the lower branch of the entropy spectrum $E_{\theta} (k)$ is found to be closer to  $1.4$. The scaling exponent is independent of  $Ta$ and $r$. The range of $k$, where this scaling holds, shifts towards higher $k$ values with increase in $Ta$. The behavior is almost similar for $Pr = 0.5$. The exponent $\gamma_2$ is close to $1.3$ for $Pr = 0.5$. The range of wave numbers for the scaling of the entropy is listed in Table~\ref{table1}. The similar scaling exponent is observed in experiments  in the frequency space instead of the  wave number space. The power spectrum $E_{\theta} (f)$ measured for the temperature field by Wu et al.~\cite{wu_etal_1990} in helium gas and Cioni et al.~\cite{cioni_etal_epl_1995} in water predict the similar exponent in frequency space.  Niemela et al.~\cite{niemela_etal_2000}, Shang and Xia~\cite{shang_xia_2001}, and Zhou and Xia~\cite{zhou_xia_2001} also found the exponent near $-7/5$ in experiments for higher values of $r$ but in the frequency space. The similar exponent for the entropy and the energy spectra in frequency space have also been observed in numerical simulations~\cite{camussi_verzicco_2004}. The sweeping time arguments imply that the scaling is similar in the wave number space. It is interesting to note that in this range of reduced Rayleigh numbers (see Fig.~\ref{nu_r}), where our scaling exponent for the spectrum is independent of the rotation speed and the Prandtl number, the exponent for the scaling of the Nusselt number is also independent of Taylor and Prandtl numbers. Figure~\ref{nu_r} shows this clearly that scaling exponent is almost independent of $Ta$ and $Pr$ for $r \ge r_t$. The critical value $r_t$, above which the universal scaling~\cite{king_etal_nature_2009, schmitz_tilgner_2010, pharasi_etal_pre_2011} $Nu \sim r^{2/7}$ holds,  decreases with increasing $Pr$ and decreasing $Ta$. It is more sensitive on $Ta$ and $Pr$ for $Pr < 0.67$, when the onset of convection is oscillatory. The upper branch of the entropy spectrum also shows scaling trend. The scaling exponent for the upper branch of the spectrum $\gamma_1$ varies between $2$ and $3$. The value of $\gamma_1$ decreases with increase in $Ta$. The bi-splitting point and the range of scaling shift towards higher $k$ at higher values of either $Ta$ or $Pr$ or both. The bi-splitting is more pronounced at higher values of $Ta$.

The effect of rotation on the energy spectrum is shown in Fig.~\ref{energy_spectrum1} at higher values of $r$ ($> 1.5 \times 10^3$). The left column of Fig.~\ref{energy_spectrum1} is for $Pr = 0.1$. The scaling quality is much better for the energy spectrum.  We observe that the energy spectrum also shows universal scaling $E_v(k) \sim k^{-\alpha}$. The exponent $\alpha$ is close to $1.4 \pm 0.1$,  which is same as the scaling exponent $\gamma_2$ of the lower part of the entropy spectrum. The range of wave numbers for the scaling of energy spectrum is is listed in Table~\ref{table1}. This is quite unexpected value and a new observation for the entropy spectrum in the wave number space obtained directly from numerical simulations. This is the result that replaces $-5/3$ if the advection of small eddies by large eddies (sweeping) dominates the dynamics. It is obtained by Kraichnan's direct interaction approximation (DIA)~\cite{kraichnan_1959} and from a very different standpoint by Mou and Weichman~\cite{mou_weichman_1993}. All the earlier exponents for the spectra were obtained in the frequency space and then indirectly connected to the exponents in the wave number space. However, there is no direct numerical or experimental evidence that the exponents for the entropy spectrum (in the wave number space) must be identical to the  exponents of the power spectrum (in the frequency space) of the temperature field.  The scaling is not so clear in the absence of rotation ($Ta = 0$), but becomes very clear for higher rotation rates.  The energy spectrum for $Pr = 0.5$ is shown in the right column of Fig.~\ref{energy_spectrum1}. The value of the exponent $\alpha$ is marginally higher for $Pr = 0.5$. The range of the universal scaling for both the  spectra shifts towards higher values of $k$ with increase in $Ta$.

The entropy spectrum for lower values of $r$ ($= 1.0\times 10^2$) is  displayed in  Fig.~\ref{entropy_spectrum2}. The bi-splitting of the entropy spectrum remains intact even at lower values of $r$. The scaling exponent $\gamma_2$ corresponding to the lower branch of the entropy spectrum  varies between  $1.8$ to $2.4$. The value of the exponent $\gamma_2$ decreases slightly with increase in $Ta$. The maximum wave number of the scaling region is in agreement with the  Bolgiano cut-off wave number (see Table~\ref{table1}). The scaling exponent corresponding to the upper branch of the entropy spectrum shows larger variation with $Ta$ for smaller values of $r$.   The scaling exponent $\gamma_1$ in this case varies between $2.2$ and $3.8$. The scaling range and the bi-splitting points shift towards higher values of $k$ at higher rotation rates, as observed in the case of higher values of $r$.  The scaling exponent for the $E_v(k)$ has relatively larger error for $r < 10^3$, where the effect of rotation is significant. Figure~\ref{energy_spectrum2} shows $E_v(k)$ for different values of $Ta$. The scaling exponent $\alpha$ for the energy spectrum is again similar to that obtained for the lower branch of the entropy spectrum. The cutoff wave number of the scaling is below $k_B$ for lower values of $r$ ($ < 5.0 \times 10^2$), but extends much beyond $k_B$ for higher values of $r$. Its value of $\alpha$ varies between  $1.7$ and $2.4$. The exponent $\alpha$ becomes smaller as the effect of rotation becomes significant. The energy and the entropy spectra for $Pr = 0.1$ and $Ta = 10^8$ are displayed in Fig.~\ref{spectrum_high_T}. The values of $Ro$ are  $1.34$ and $2.99$ for the data points in the left and right columns respectively. The upper row displays the entropy spectrum for (a) $r = 1.0 \times 10^2$ and (b) $r = 5.0 \times 10^2$. The bi-splitting of the entropy spectrum is observed to be less pronounced for $r = 1.0 \times 10^2$ compared to that observed at $r =5.0 \times 10^2$. The scaling exponent $\gamma_1$ varies from $2.16$ and $2.8$, while $\gamma_2$ varies from $1.89$ to $2.6$. The energy spectrum (the lower row of Fig.~\ref{spectrum_high_T}) for (c) $r = 1.0 \times 10^2$ and (d) $r = 5.0 \times 10^2$.  The exponent of the energy spectrum $\alpha$ is found to vary from $1.5$ to $2.0$. The maximum error margins to the exponents is  $14 \%$ in this case. 

Figure~\ref{exonents} show the variation of scaling exponents $\gamma_1$, $\gamma_2$ and $\alpha$ with $Ta$ for different values of $r$. The exponent $\gamma_1$  decreases with increasing $Ta$, while $\gamma_2$ remains independent of $Ta$ [Fig.~\ref{exonents} (a) and (b)]. However the values of $\gamma_1$ and $\gamma_2$ at a given value of $Pr$ are larger at smaller values of $r$. The exponent $\alpha$ is apparently independent of $Ta$. Its value for a given $Pr$ is also larger for smaller values of $r$ [see, Fig.~\ref{exonents} (c)]. 
 
\section{Discussions}
We now begin by outlining the reason behind the observations by first recalling the Kolmogorov argument (K41) for the situation without any convection. The energy budget of the unforced Navier-Stokes equation, with  variables maintaining their dimensions, is known to be 

\begin{equation}
\epsilon_K = \frac{dK}{dt}=-2 \nu \int \left( \frac{\partial{v_{i}}}{\partial{x_{j}}}\right)^2 dV,
\end{equation}
\noindent
where $K = \int (v^2/2) dV$ is the total kinetic energy. This shows that $K$ is conserved in the inviscid limit and now if we inject energy at the rate $\epsilon_K = dK/dt$ into the system at large length scales, then a stationary state can be achieved wherein  energy is pumped into the system at large length scales and dissipated at short length scales by molecular viscosity. The contribution due to the nonlinear terms integrate out to zero  in the total energy budget, and the energy transfer from one scale to another occurs at the constant rate $\epsilon_K$. This intermediate scale is the inertial range of Kolmogorov and in this range all physical quantities are determined by only two quantities, the scale itself ($l$ in coordinate space and $k$ in momentum space) and $\epsilon_K$. Dimensional analysis now leads to the well known Kolmogorov energy spectrum. 

\begin{equation}
E_v(k) = C\epsilon_K^{2/3}k^{-5/3},\label{kolmogorov_spectrum}
\end{equation}
where $K=\int E_v(k)dk$ and C is  a numerical constant. We address the questions: what is the analogue of the above argument for the Rayleigh-B\'{e}nard convective turbulence, and how uniform rotation of the fluid layer about a vertical axis affects it. We also have convective entropy $\Phi = \int ({\theta}^2/2) dV$ in the case of RB convection, where $\theta$ stands for the change in the temperature field due to convection. The total entropy may also be written as:  $\Phi = \int E_{\theta}(k)dk$, where $E_{\theta}(k)$ gives the entropy spectrum. The rate of change the convective entropy $\epsilon_{\Phi}$ is:

\begin{equation}
\epsilon_{\Phi}=\frac{d\Phi}{dt}=-\kappa \int \left( \frac{\partial{\theta}}{\partial{x_{i}}}\right)^2 dV.
\end{equation}

We see from Eqs.(\ref{momentum})-(\ref{continuity}) in dimensional form that the rate of change of the kinetic energy $K$ and the entropy $\Phi$ are:

\begin{equation}
\frac{dK}{dt} = \alpha g \int (v_3 \theta)dV - 2 \nu \int \left( {\frac{\partial v_i}{\partial x_j}} \right)^2 dV,
\end{equation}

\begin{equation}
\frac{d\Phi}{dt} = - \kappa \int \left( {\frac{\partial \theta}{\partial x_i}} \right)^2 dV \nonumber \\
+  \frac{({\Delta T})}{d} \int (v_3 \theta) dV.
\end{equation}

This approach assumes that there is at least a close to linear velocity profile. Defining the dimensionless temperature $\tilde{\theta} = \theta/({\Delta T})$ and the corresponding total entropy $\tilde{\Phi} = \int ({\tilde{\theta}}^2/2) dV$, appropriate subtraction leads to 

\begin{eqnarray}
\frac{d}{dt}[K &-& \alpha ({\Delta T}) d g  \tilde{\Phi}] = - 2 \nu \int \left( \frac{\partial v_i}{\partial x_j} \right)^2 dV \nonumber \\ 
&+&\alpha({\Delta T}) d g \kappa \int \left( \frac{\partial \tilde{\theta}}{\partial x_i} \right)^2 dV.
\label{eq_mixed}
\end{eqnarray}

Clearly the conserved quantity in the inviscid limit is  $K -  \alpha ({\Delta T})dg  \tilde{\Phi}$. The Kolmogorov kind of argument that we have given above works only in the two limits:\\
(A)  $K >>  \alpha ({\Delta T}) d g \tilde{\Phi}$ gives K41-like behavior, and\\
(B)  $K <<  \alpha ({\Delta T}) d g \tilde{\Phi}$ yields BO-like  behavior.\\
It is pertinent to ask when would one flux dominate the other. The analysis~\cite{castaing_1989} of the Nusselt number which leads to the exponent of $2/7$ allows us to infer that $K \propto Ra^{\gamma}$ and $\phi(\delta T) \propto Ra^{2 \gamma-1}$ where the exponent $\gamma$ is $6/7$. This tells us that $K$ will dominate at large $Ra$ and that should be the range where Kolmogorov like spectrum should hold. This is in conformity with the results of Fig~\ref{entropy_spectrum1}.

We have K41-like behavior if the RHS of Eq.~\ref{eq_mixed} is negative. We have a BO-like situation if the RHS is positive. The LHS of Eq.~\ref{eq_mixed} changes sign if the rate of change of $K -  \alpha ({\Delta T}) d g \tilde{\Phi}$ changes sign. The RHS, in addition to the gradients of the temperature and velocity fields, depends on the Prandtl number. This is apparent if one takes $\nu$ or $\kappa$ common to both terms of the RHS of Eq.~\ref{eq_mixed}. Consequently the spectra are sensitive to the Prandtl numbers. The scaling is unclear when both the terms on RHS are comparable. This seems to happen for $Pr = 7$ case of Mishra and Verma~\cite{mishra_etal_pre_2010}. Similarly for very low $Pr$, Eq.~\ref{eq_mixed} indicates that even if $K$ were to dominate on LHS, there could be an injection of energy at short scales due to $\tilde{\Phi}$. Conventional wisdom says that there is a length scale which demarcates between K41-like and BO-like regimes. This is  $L_B/d = (\overline{Nu})^{1/2}/(Ra Pr)^{1/4}$. It is BO-like if $l > L_B$ and K41-like on the other side. We want to argue that this simplistic. 

Returning to Eq.~\ref{eq_mixed}, we see that a Kolmogorov type argument that led to the well known result (Eq.~\ref{kolmogorov_spectrum}) will in the case of convective turbulence need the flux of the combined quantity $K -  \alpha ({\Delta T}) g d \tilde{\Phi}$. If $\epsilon_K$ is the flux of the kinetic energy $K$ (pure Kolmogorov) and if $\epsilon_{\tilde{\Phi}}$ is the flux of the entropy (pure Bolgiano), then for the total kinetic energy flux the dimensional argument gives the energy spectrum
\begin{equation}
E(k)= C [\epsilon_K -  \alpha ({\Delta T}) g d \epsilon_{\tilde{\Phi}}]^{2/3} k^{-5/3} \label{eq1}
\end{equation}
$\epsilon_K$ is scale independent but  $\epsilon_{\tilde{\Phi}}$ is not. If $\epsilon_{\tilde{\Phi}}$ is to be determined by  $\epsilon_K$ and $k$, then clearly $\epsilon_{\tilde{\Phi}}$  =  $D \epsilon_K^{1/3} k^{2/3}$ with $D$ a constant.  Eq.~\ref{eq1} then becomes
\begin{equation}
E(k)= C \epsilon_K^{2/3}\left[ 1- D  \alpha ({\Delta T}) d g k^{2/3} \epsilon_K^{-2/3} \right]^{2/3}k^{-5/3}\label{eq2}
\end{equation}
The part within [...] can be written as $1-({k}/{k_1})^{2/3}$and it is clear that K41-like situation can hold only if $k < k_1$ and on the upper end it is bounded by $2 \pi L_B^{-1}$. In the range where it is valid, it will show  an effective exponent $\alpha_1$ which is clearly greater than  $5/3$ and is given by\\
\begin{equation}
\alpha_1  = -\left[ 5/3 + \frac{4}{9}\frac{(k/k_1)^{2/3}}{[1-{(k/k_1)^{2/3}}]} \right]. \label{eq3}
\end{equation}
The effective exponent shown above will allow an approximate scaling behavior only if the function on the RHS of Eq.~\ref{eq3} varies slowly with k.

Now to the situation where the thermal flux dominates and we are in a BO-like situation. Dimensional argument now gives
\begin{equation}
E(k)=C^{\prime}\left[\epsilon_{\tilde{\Phi}}-\frac{\epsilon_K}{\alpha ({\Delta T}) dg} \right]^{2/5}\left(\alpha g \right)^{4/5} k^{-11/5}
\end{equation}
In this domain $\epsilon_{\tilde{\Phi}}$ is scale independent but $\epsilon_K$ is not. In fact dimensional analysis  gives 
\begin{equation}
\epsilon_K = D^{\prime}\epsilon_{\tilde{\Phi}}^{3/5}(\alpha g)^{6/5}k^{-4/5}\label{eq4}
\end{equation}
With Eq.~\ref{eq4} in mind, we write
\begin{equation}
E(k)= C'{\epsilon_{\tilde{\Phi}}}^{2/5}\left[1-(k_2/k)^{4/5}\right]^{2/5} (\alpha g)^{4/5} k^{-11/5}\label{eq5}
\end{equation} 
The scale $k_2$ is given by $k_2^{4/5} = D^{\prime}(\alpha g)^{1/5}               \epsilon_{\tilde{\Phi}}^{-2/5} d (T_1- T_2)$ and BO is valid only if $k> k_2$. On the lower end it is bounded by $2 \pi L_B^{-1}$.
The effective exponent $\alpha_2$ follows  as 
\begin{equation}
\alpha_2   = -\left[\frac{11}{5}-\frac{8}{25}\frac{{(k_2/k)}^{0.8}}{[1-({k_2}/{k})^{0.8}]}\right]\label{eq6}
\end{equation}
The absolute value of the effective exponent now is smaller than  11/5 and is  $k$ dependent which makes a long scaling range difficult to obtain. Only if one is in a region where the function above is not changing fast that  we can see the appearance of a definite exponent. Note that by changing the Rayleigh number as one crosses from K41 to BO, there can be significant corrections on both sides and large unexpected changes in the exponent may occur. In particular a Bolgiano region can show an exponent much smaller than 2.2. Note also that rotation suppresses the energy flux relative to the thermal flux and hence can aid a transition to Bolgiano.

\section{Conclusions}
We have presented the results for the entropy and the energy spectra computed from DNS with high accuracy for turbulent flows in low-Prandtl-number Rayleigh-B\'{e}nard convection with uniform rotation about a vertical axis. The entropy spectrum shows bi-splitting in wave number space. The scaling exponent for the upper branch of the entropy spectrum is not universal. The scaling exponent for the lower branch of the entropy spectrum  scales with wave number as $k^{-1.4 \pm 0.1}$ for larger values of the reduced Rayleigh number. The energy spectrum also shows similar scaling and scales with $k$ as $k^{-1.5 \pm 0.1}$. The scaling is found to be universal for higher values of $r$ ($> 10^3$). It is observed for wave numbers below a cut-off wave number $k_B$ corresponding to the Bolgiano length. For smaller values of the reduced Rayleigh number ($r < 10^3$), the scaling exponent varies between $-1.7$ and $-2.4$.  The presence of uniform rotation appears to make BO scaling more accessible. 

\begin{center}
\bf{ACKNOWLEDGMENTS}
\end{center}
We have benefited from fruitful discussions with Priyanka Maity, Arnab Basak, and Rohit Raveendran.


\begin{thebibliography}{100}
\bibitem{siggia_1994}
{E.D. Siggia}, {Annu. Rev. Fluid Mech.} {\bf 26}, {137} {(1994)}.
\bibitem{ahlers_review}
{G. Ahlers, S. Grossmann, and D. Lohse}, {Rev. Mod. Phys.} {\bf81}, {503} {(2009)}.
\bibitem{lohse_xia_2010} 
{D. Lohse and K.-Q. Xia}, {Annu. Rev. Fluid Mech.} {\bf 42}, {335} {(2010)}.
\bibitem{kerr_1996}
{R.M. Kerr}, {J. Fluid Mech.} {\bf 310}, {139} {(1996)}.
\bibitem{cioni_etal_1997}
{S. Cioni, S. Ciliberto, and J. Sommeria}, {J. Fluid Mech.} {\bf 335}, {111} {(1997)}.
\bibitem{heslot_1987}
{F. Heslot, B. Castaing, A. Libchaber}, {Phys. Rev. A} {\bf 36}, {5870} {(1987)}.
\bibitem{castaing_1989}
{B. Castaing, G. Gunaratne, F. Heslot, L. Kadanoff, A. Libchaber}, {J. Fluid Mech.} {\bf 204}, {1} {(1989)}.
\bibitem{wu_etal_1990}
{X.-Z. Wu,  L. Kadanoff, A. Libchaber, and M. Sano}, {Phys. Rev. Lett.} {\bf 64}, {2140} {(1990)}.
\bibitem{niemela_etal_2000}
{J.J. Niemela, L. Skrbek, K.R. Sreenivasan, and R.J. Donnelly}, {Nature}      {\bf 404}, {837} {(2000)}.
\bibitem{grossmann_lohse_2000}
{S. Grossmann and D. Lohse}, {J. Fluid Mech.} {\bf 407}, {27} {(2000)}.
\bibitem{kadanoff_2001}
{L.P. Kadanoff}, {Physics Today} {\bf 54(8)}, {34} {(2001)}.
\bibitem{fauve_epl_2003}
{S. Auma\^{i}tre and S. Fauve}, {Europhys. Lett.} {\bf 62 (6)}, {822} {(2003)}.
\bibitem{niemela_etal_1986}
{J.J. Niemela and R.J. Donnelly}, {Phys. Rev. Lett.} {\bf 57}, 2524 (1986).    
\bibitem{julien_etal_1996}
{K. Julien, S. Legg, J. McWilliams, and J. Werne}, {J. Fluid Mech.} {\bf 322}, {243} {(1996)}.
\bibitem{liu_ecke_1997}
{Y. Liu and R.E. Ecke}, {Phys. Rev. Lett.} {\bf 79}, {2257} {(1997)}.
\bibitem{kunnen_etal_epl_2008}
{R.P.J. Kunnen, H.J.H. Clercx and B.J.Geurts}, {Europhys. Lett.} {\bf 84}, {24001} {(2008)}.
\bibitem{stevens_etal}
R.J.A.M. Stevens, J.-Q. Zhong, H.J.H. Clercx, G. Ahlers, and D. Lohse, {Phys. Rev. Lett.} {\bf 103}, 024503 (2009).
\bibitem{king_etal_nature_2009}
{E.M. King, S. Stellmach, J. Noir, U. Hansen, and J.M. Aurnou}, {Nature} {\bf 457}, {301} {(2009)}.
\bibitem{schmitz_tilgner_2010}
{S. Schmitz and A. Tilgner}, {Geophys. Astrophys. Fluid Dynamics} {\bf 104}, 481 (2010).
\bibitem{pharasi_etal_pre_2011}
{H.K. Pharasi, R. Kannan,  K. Kumar, and J.K. Bhattacharjee}, {Phys. Rev. E}     {\bf 84}, {047301} {(2011)}.
\bibitem{stevens_2013}
{R.J.A.M. Stevens, H.J.H. Clercx, and D. Lohse}, {Euro. J. Mech. B/Fluids} {\bf 40}, {41} {(2013)}.
\bibitem{K41}
{A.N. Kolmogorov}, {Dokl. Akad. Nauk. SSSR} {\bf 30}, {299} {(1941)}.
\bibitem{bolgiano_1959}
{R. Bolgiano}, {J. Geophys. Res.} {\bf 64}, {2226} {(1959)}.
\bibitem{obukhov_1959}
{A.M. Obukhov}, {Dokl. Akad. Nauk. SSSR} {\bf 125}, {1246} {(1959)}.
\bibitem{ashkenazi_steinberg_1999}
{S. Ashkenazi and V. Steinberg}, {Phys. Rev. Lett.} {\bf 83}, {4760} {(1999)}.
\bibitem{shang_xia_2001}
{X.-D. Shang and K.-Q. Xia }, {Phys. Rev. E}{\bf 64}, {065301(R)} {(2001)}.
\bibitem{zhou_xia_2001}
{S.-Q. Zhou  and K.-Q. Xia }, {Phys. Rev. Lett.} { \bf 87}, {064501} {(2001)}.
\bibitem{calzavarini_etal_2002}
{E. Calzavarini, F. Toschi, and R. Tripiccione}, {Phys. Rev E} {\bf 66} {016304} (2002).
\bibitem{kunnen_etal_pre_2008}
{R.P.J. Kunnen, H.J.H. Clercx, B.J.Geurts, L.J.A. van Bokhoven, R.A.D. Akkermans, and R. Verzicco}, {Phys. Rev. E} {\bf 77}, {016302} {(2008)}.
\bibitem{mishra_etal_pre_2010} 
{P.K. Mishra and M.K. Verma}, {Phys. Rev. E} {\bf 81}, {056316} {(2010)}.
\bibitem{kraichnan_1959}
{R.H. Kraichnan}, {J. Fluid Mech.} {\bf 5}, {497} {(1959)}.
\bibitem{l'vov_1991}
{V. S. L'vov}, {Phys. Rev. Lett.} {\bf 67}, {687} {(1991)}.
\bibitem{l'vov_1992}
{V.S. L'vov  and G.E. Falkovich}, {Physica D} {\bf 57}, {85} {(1992)}.
\bibitem{brandenburg_1992}
{A. Brandenburg}, {Phys. Rev. Lett.} {\bf 69}, {605} {(1992)}.
\bibitem{mou_weichman_1993}
{C.Y. Mou and P.B. Weichman}, {Phys. Rev. Lett.} {\bf 70}, {1101} {(1993)}.
\bibitem{chandrasekhar:book_1961} 
{S. Chandrasekhar}, {\em Hydrodynamic and Hydromagnetic Stability}, {Oxford University Press, Oxford (1961)}, reprinted by Dover (New York, 1981).
\bibitem{kaneda_etal_2003} 
{Y. Kaneda, T. Ishihara, M. Yokokawa, K. Itakura, and A. Uno} {Phys. Fluids} {\bf 15} {L 21} (2003).
\bibitem{vincent_1999}
{A.P. Vincent and D.A. Yuen}, {Phys. Rev. E} {\bf 60}, {2957} (1999).
\bibitem{tarang}
{M.K. Verma, A. Chatterrjee, K. S. Reddy, R. K. Yadav, S. Paul, M. Chandra, and R. Samtaney}, {Pramana} {\bf 81}, {617} (2013).
\bibitem{cioni_etal_epl_1995}
{S. Cioni, S. Ciliberto, and J. Sommeria}, {Europhys. Lett.} {\bf 32 (5)}, {413} {(1995)}.
\bibitem{camussi_verzicco_2004}
{R. Camussi and R. Verzicco}, {Eur. J. Mech. B/Fluids} {\bf 23}, 427 (2004).
\end{thebibliography}
\end{document}